\documentclass[aps,prl, twocolumn]{revtex4-1}
\usepackage{amsmath, amssymb, graphicx, color}
\def\aap{Astron. Astrophys.}
\def\aj{Astron.J.}
\def\aveh#1{#1_h}
\def\avel#1{#1_L}
\def\chiarm{\chi_{\rm arm}}
\def\phih{\aveh{\phi}}
\def\chih{\aveh{\chi}}
\def\chil{\avel{\chi}}
\def\zetah{\aveh{\zeta}}
\def\zetal{\avel{\zeta}}
\def\chihi{\chi_{h,i}}
\def\hpre{\mathcal{H}_{\rm pre}}

\def\erf{\mathrm{erf}}
\def\hmpc{\mathcal{H}_8}
\def\lmpc{L_8}
\def\kmpc{k_8}

\begin{document}
\title{High-redshift Mini-haloes from Modulated Preheating}
\author{Zhiqi Huang}
\affiliation{School of Physics and Astronomy, Sun Yat-sen University, 2 Daxue Road, Zhuhai, CHINA}
\email{huangzhq25@mail.sysu.edu.cn}

\date{\today}

\begin{abstract}

Intermittent type of primordial non-Gaussian fluctuations from modulated preheating can produce an overabundance of $\sim 10^8M_\odot$ mini-haloes at high redshift $z\gtrsim 20$. This may have a significant impact on the formation of high-redshift supermassive black holes.
  
 \end{abstract}

\maketitle

\section{Introduction \label{sec:intro}}

Preheating is a postulated nonlinear process following the end of early-universe inflation, where the inflaton field quickly dumps its energy into other field(s) via parametric resonance. In the separate-universe view, preheating happens almost independently in each Hubble patch, whose typical comoving size is $\sim 25$ orders of magnitude below cosmological scales. Refs.~\cite{Suyama07, Chambers08} suggested that, however,  preheating may still leave some imprint on the large-scale structure of the universe, but could not convincingly justify their statement due to insufficient numeric accuracies in their calculation. Ref.~\cite{BFHK09} (hereafter BFHK) significantly improved the numeric tools and first explicitly demonstrated such an effect. The preheating dynamics is closely tied to the background field values averaged in each Hubble patch. The background trajectory of inflaton and other coupled field(s) is usually chaotic and is very sensitive to the initial conditions for preheating, i.e., field fluctuations at the end of inflation, which, as the standard cosmic inflation story tells, can be correlated on cosmological scales. The inflaton field itself serves as the clock for the separate universe during inflation. Thus the fluctuations of inflaton at the end of inflation can be gauged away and do not produce any physical modulation on the preheating dynamics. The other field(s), which we dub modulator(s), may generate curvature fluctuations on cosmological scales by modulating the averaged equation of state in each Hubble patch. This effect opens an exciting window to the rich physics at the end of inflation.

BFHK first pointed out that modulated preheating may explain the anomalous cold spot in the cosmic microwave background (CMB) radiation ~\cite{ColdSpot1, ColdSpot2}. They studied a two-field model with Lagrangian density
\begin{equation}
\mathcal{L}(\phi,\chi) = \frac{1}{2}\partial^\mu\phi\partial_\mu\phi +  \frac{1}{2}\partial^\mu\chi\partial_\mu\chi- \frac{1}{4}\lambda\phi^4 - \frac{1}{2}g^2\phi^2\chi^2, \label{eq:L}
\end{equation}
where $\phi$ is the inflaton and $\chi$ the modulator. The dynamics of the background trajectory, $\left(\phih(t), \chih(t)\right)$, can be described as a billiard ball rolling back and forth in a spindle-shaped potential well. (See Fig.~2 of BFHK. Here a subscript $h$, unless otherwise stated, represents spatial average within the Hubble patch.) If the billiard ball enters one of the arms of the spindle shape, where $\left\vert\chih\right\vert \gg\left\vert\phih\right\vert$, the $\phi$ field becomes heavy and its harmonic-like oscillations slightly pull down  the effective equation of state, $w_{\rm eff} \equiv \aveh{\rho}/\aveh{p}$, where $\rho$ and $p$ are respectively the total energy density and the total pressure. Therefore, for Hubble patches where the background trajectory enters the spindle arms, the scale factor of the Hubble patch, $a \propto \aveh{\rho}^{-\frac{1}{3(1+w_{\rm eff})}}$, grows faster as the average energy density drops, relatively to other Hubble patches where the background trajectory does not enter the spindle arms. The probability that the background trajectory enters the spindle arms is modulated by the initial $\chih$ value at the end of inflation (hereafter denoted as $\chihi$), which is then modulated by the long wavelength $\chi$ fluctuations on cosmological scales. In this way, the superhorizon $\chi$ fluctuations prepared by inflation modulate the number of expansion e-folds, which can be translated to the comoving curvature fluctuations $\zeta$ with the $\delta N$ formula~\cite{SB90, SS96, Tanaka03, KLV10, Suyama13, IMR18}. Such a non-Gaussianity in $\zeta$, by its nature of origin, is  usually {\it intermittent in configuration space}. The anomalous CMB cold spot can be interpreted as a spatial region on the last scattering surface where the $\chih$ value enhances the probability of background trajectory entering the spindle arms.

In the model studied in BFHK, and in many other preheating models, $\chih$ grows exponentially, $\chih(t) \sim e^{\mu t}$ via parametric resonance during the linear regime of preheating,  where $\mu$ is the Floquet exponent and $t$ the cosmological time. Therefore, as shown in Fig.~1 of BFHK, if a particular initial $\chihi = \chiarm$ enhances the probability for the background trajectory entering the spindle arms, an initial $\chihi = e^{n\mu T}\chiarm$, where $n$ is an integer and $T$ the period of inflaton oscillations during the linear regime, tends to enhance the probability of background trajectory entering the arms, too.

The $\phi^4$ inflation model predicts a tensor to scalar ratio $r\gtrsim 0.2$, which is ruled out by recent CMB observations~\cite{PlanckInflation15}. Replacing $\phi^4$ term with a viable inflation model, however, is not likely to be a big problem for us to study modulated preheating, since here the only role inflation plays is to prepare initial conditions for preheating. Similarly, the bottom of the potential well does not have to be exactly spindle shaped with two long arms. The variety of possibilities makes it difficult to confront modulated preheating models with observations. Thus, it is important to extract common features of modulated preheating models and construct a parameterization as model-independent as possible. This is the subject of the next section.

Throughout this article we use natural units $c=\hbar=1$ and assume a cosmology with the amplitude of primordial scalar power spectrum $A_s = 2.14\times 10^{-9}$, the spectral index $n_s = 0.965$,  the baryon density $\Omega_b h^2=0.0221$, cold dark matter density $\Omega_ch^2 = 0.119$, and the reduced Hubble constant $h=0.677$.  

\section{Model  \label{sec:model}}

We consider a simplified scenario with only one modulator field $\chi$ whose cosmic ensemble average is zero, and assume equal-amplitude non-Gaussian $\zeta$ spikes triggered by a series of log-uniform initial $\chihi = \pm e^{n\mu T}\chiarm$ $(n=0,\pm 1, \pm 2,\ldots)$. Hereafter, if no possible confusion arises, we will drop the subscript $i$ for readability. Written explicitly, the averaged $\zeta$ in a Hubble patch is given by
\begin{equation}
  \zetah = \sum_{n=-\infty}^\infty A_\zeta \, \delta_D\left(\ln \frac{\left\vert\chih\right\vert}{\chiarm} - nW\right), \label{eq:zeta}
\end{equation}
where $\delta_D$ is the Dirac delta function. The $W$ parameter is the Floquet exponent in unit of the period of inflaton oscillation, and the $A_\zeta$ parameter represents the amplitude of modulation. 

Indeed, the parameterization in Eq.~\eqref{eq:zeta} only describes an idealized scenario. BFHK has shown, by running high-precision lattice simulations for the concrete model in Eq.~\eqref{eq:L}, that due to the contribution from the subhorizon modes of $\phi$ and $\chi$ fluctuations, $A_\zeta$ can have a weak $n$ dependence. Moreover, as shown in Fig.~1 of BFHK, there can be more than one log-uniform series of $\zeta$ spikes with different amplitudes. Nevertheless, the model defined by Eq.~\eqref{eq:zeta} is a basic building block of general features from modulated preheating. The results presented in this work should be understood as qualitative and general estimations for modulated preheating models, rather than precise calculations for a specific model.

Eq.~\eqref{eq:zeta} gives the mapping from $\chi$ to $\zeta$ on a scale $\sim \hpre^{-1}$, where $\hpre$ denotes the Hubble parameter during preheating. For the purpose of studying cosmology, we need to work out the mapping from $\chi$ to $\zeta$ on much larger scales.

Let us consider a cosmological-size comoving volume $\sim L^3$, where $L \gg \hpre^{-1}$ is a scale relevant for cosmological observations, typically within a few orders of magnitude from $\mathrm{Mpc}$. The spatially averaged $\zeta$ in this volume, denoted by $\zetal$, can be written as
\begin{equation}
  \zetal = \int \aveh{\zeta}(\chih) P(\chih; \chil) d\chih, \label{eq:zL}
\end{equation}
where $P(\chih;\chil)$ is the conditional probability density function of $\chih$ for a given $\chil$ (averaged $\chi$ in the $\sim L^3$ volume). For simplicity we assume $\chi$ fluctuations follow Gaussian statistics,
\begin{equation}
  P(\chih; \chil) = \frac{1}{\sqrt{2\pi}\sigma_L}e^{-\frac{(\chih-\chil)^2}{2\sigma_L^2}}\, , \label{eq:P}
\end{equation}
where $\sigma_L$ can be computed from  the $\chi$-field power spectrum $P_\chi(k)$,
\begin{equation}
  \sigma_L^2 = \int^{\mathcal{H}}_{1/L} \frac{k^3P_\chi(k)}{2\pi^2} \frac{dk}{k}. \label{eq:sigma2}
\end{equation}

We choose the pivot scale to be $l_{\rm pivot} = 8 h^{-1}\mathrm{Mpc}$ and expand the power spectrum
\begin{equation}
  \frac{k^3P_\chi(k)}{2\pi^2} \sim e^{c_0+c_1\ln k_8 + \frac{1}{2}c_2 (\ln k_8)^2+\ldots}, \label{eq:expand}
\end{equation}
where $\kmpc \equiv k l_{\rm pivot}$. Well established observational constraints on quasi-linear scales $\gtrsim 8 h^{-1}$ favor a simple inflation model, where the inflaton was light when its fluctuations on scales $\gtrsim l_{\rm pivot}$ were generated. For simplicity we assume the same scenario for the modulator $\chi$ field. The power spectrum of light field fluctuations $\sim\left(\frac{H}{2\pi}\right)^2$, $H$ being the Hubble expansion rate, is nearly scale invariant. Thus, we approximate $c_1\approx 0$, truncate at the next leading order, and re-parameterize Eq.~\eqref{eq:expand} as 
\begin{equation}
  \frac{k^3P_\chi(k)}{2\pi^2} =  \chiarm^2  e^{2\lambda W -\frac{1}{2}\alpha \left(\ln \kmpc\right)^2}, \label{eq:Pchi}
\end{equation}
where $\alpha \equiv -c_2$ and $\lambda \equiv \frac{c_0- 2\ln \chiarm}{2W}$ are constants. The advantage of using $\lambda$ instead of $c_0$ is that the self-similarity of $\chih$ in Eq.~\eqref{eq:zeta} makes the mapping from $\chi_L$ to $\zeta_L$ periodic in $\lambda$ with period=$1$. If the $\chi$ field continues to be light until the end of inflation, one would find $\alpha\approx 0$. However, a positive $\alpha\ll 1$ could be caused by a slowly increasing mass of $\chi$ field towards the end of inflation. We do not consider complex scenarios with tachyonic or parametric resonance instabilities, which may lead to a negative $\alpha$. In summary, hereafter we will restrict $\lambda\in[0,1)$ and $0\le \alpha\ll 1$.

Substituting Eq.~\eqref{eq:Pchi} into Eq.~\eqref{eq:sigma2} , we obtain
\begin{equation}
  \frac{\sigma_L}{\chiarm} =  \left(\frac{\pi }{2\alpha}\right)^{\frac{1}{4}}  e^{\lambda W}  \sqrt{\erf\left(\sqrt{\frac{\alpha}{2}} \ln \hmpc\right)- \erf\left(-\sqrt{\frac{\alpha}{2}}\ln \lmpc\right)}, \label{eq:sL}
\end{equation}
where $\hmpc \equiv \mathcal{H}l_{\rm pivot}$ and $\lmpc \equiv L/l_{\rm pivot}$.

Given that inflation lasts for 50-60 efolds, we have $\ln \hmpc =  55 \pm \text{a few}$. For very small $\alpha\lesssim \frac{1}{\ln^2 \hmpc} \sim 10^{-3}$, fixing $\ln \hmpc = 55$, which we will do unless otherwise specified, could introduce an error $\lesssim 5\%$, which is tolerable for the purpose of qualitative estimations. For larger $\alpha$, the exact value of $\ln \hmpc$ is more irrelevant, as $\erf\left(\sqrt{\frac{\alpha}{2}} \ln \hmpc\right)$ will be very close to $1$.

The combination of Eqs.~\eqref{eq:zeta}, \eqref{eq:zL}, \eqref{eq:P}, \eqref{eq:sL} gives a mapping from $\frac{\chi_L}{\chiarm}$ to $\frac{\zetal}{A_\zeta}$, which is parameterized by four additional parameters $W$, $\alpha$, $\lambda$, and $\lmpc$.

Having the mapping from $\chi_L$ to $\zeta_L$ worked out, we now proceed to compute cosmological observables by running N-body simulations with the modified initial conditions from modulated preheating. The basic idea is to generalize random Gaussian fluctuations of $\chi$ and map them to $\zeta$ fluctuations.

\section{Overabundance of High-redshift Mini-haloes \label{sec:halo}}

For a cosmological N-body simulation, we prepare the initial matter density fluctuation with the following procedures. 

\begin{enumerate}
\item{Realize random Gaussian fluctuations of $\chi_L/\chiarm$ in Fourier space. The smoothing scale $L=B/N$ is the simulation resolution, where $B$ is the box side length and $N^3$ is the total number of particles. }
\item{Fourier transform $\chi_L/\chiarm$ to configuration space.}  
\item{Map $\chi_L/\chiarm$ to $\zeta_L$.}
\item{Transform $\zeta_L$ to Fourier space.}
\item{Add the standard Gaussian component of $\zeta$, that is, a random Gaussian realization with the standard power spectrum defined by $A_s$ and $n_s$.}
\item{Use the linear transfer function to map the total comoving curvature fluctuations to matter density fluctuations.}
\item{Transform the matter density fluctuations back to configuration space.}  
\end{enumerate}

In step 1 we need to specify the $k=0$ mode of $\chi_L/\chiarm$ in the box, which we dub $\chi_B/\chiarm$. Eq.~\eqref{eq:Pchi} implies that, for scales not too far away from $L_{\rm pivot}$, the root mean square fluctuation of $\chi/\chi_{\rm arm}$ per e-fold is $\sim e^{\lambda W}$. Again for simplicity, we assume a vanishing cosmic background $\chi/\chi_{\rm arm}$ for the observable universe. In other words, we assume $|\chi_B/\chiarm| \ll  e^{\lambda W}$ for $B\gtrsim 10 h^{-1}\mathrm{Gpc}$. For smaller box simulations, typically $|\chi_B/\chiarm| < 2 e^{\lambda W}\sqrt{\ln\frac{10 h^{-1}\mathrm{Gpc}}{B}}$ at $2\sigma$ ($95\%$) confidence level. 

We use the N-body code COLA~\cite{COLA13} with fast integration schemes, which is sufficiently accurate for the purpose of qualitative studies and allows us to do ensemble averages of many simulations. Haloes are determined with friends-of-friends algorithm~\cite{FOF85} with a linking-length factor $0.2$.

\begin{figure}
  \includegraphics[width=0.48\textwidth]{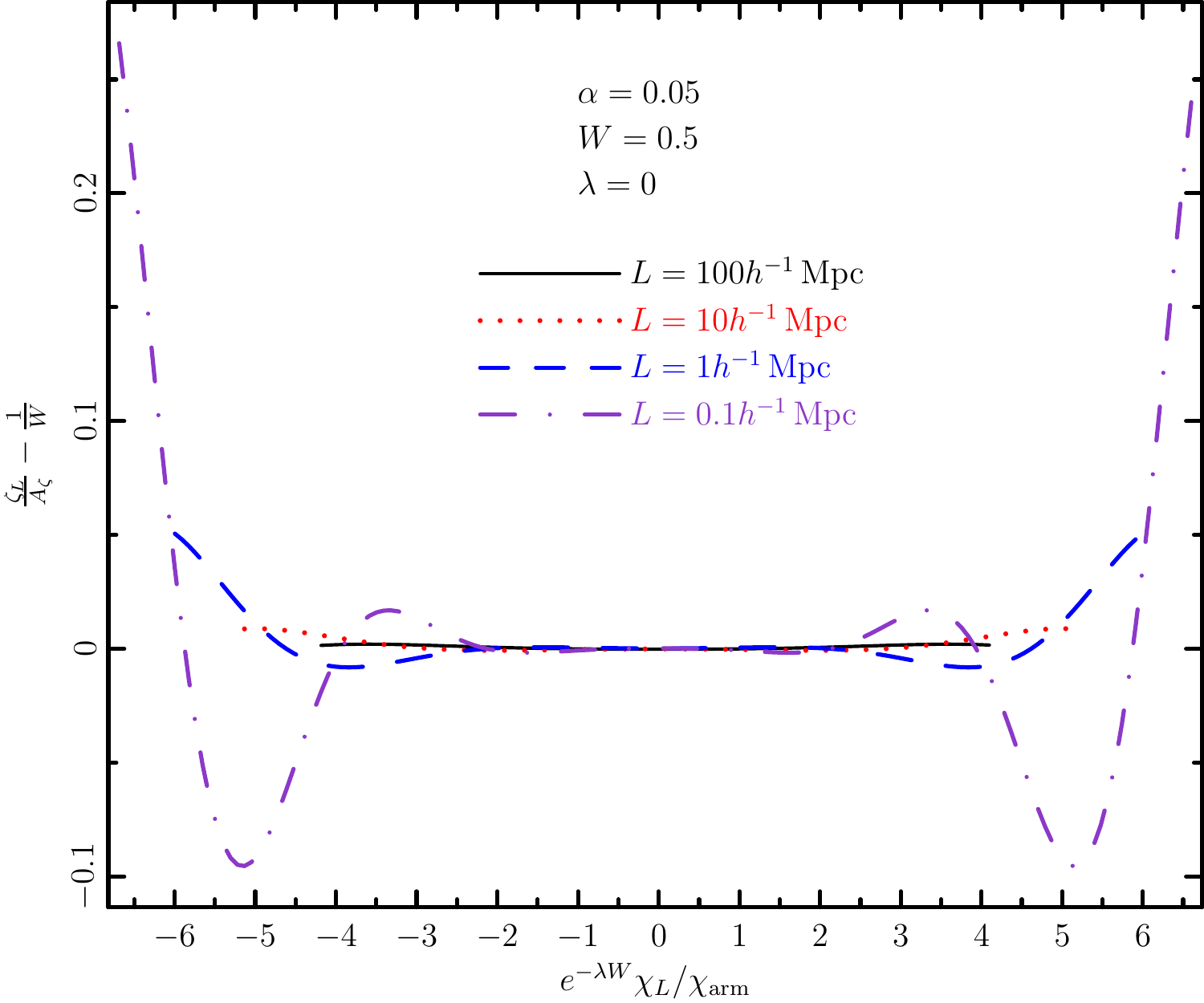}
  \caption{Mappings from $\frac{\chi_L}{\chiarm}$ to $\frac{\zeta_L}{A_\zeta}$  for $\chi_L/\chiarm$ in the interval of $2\sigma$ ($95\%$) confidence level: $|e^{-\lambda W}\frac{\chi_L}{\chiarm}|<2 \sqrt{\ln\frac{10 h^{-1}\mathrm{Gpc}}{L}}$. \label{fig:zeta}}
\end{figure}

We are interested in the models with a non-vanishing $\alpha$, in which case the very-small-scale ($L\ll \mathrm{Mpc}$) fluctuations of $\chi$ are suppressed. Figure~\ref{fig:zeta} shows the mappings from $\chi_L/\chiarm$ to $\zeta_L$ for a typical set of parameters and a hierarchy of scales: $100h^{-1}\mathrm{Mpc}$, $10h^{-1}\mathrm{Mpc}$, $1h^{-1}\mathrm{Mpc}$, and $0.1 h^{-1}\mathrm{Mpc}$. The scale $100h^{-1}\mathrm{Mpc}$ is a typical representation of linear scales, on which the primordial fluctuations are stringently constrained by CMB and large-scale structure observations. The scales $10h^{-1}\mathrm{Mpc}$, $1h^{-1}\mathrm{Mpc}$, and $0.1 h^{-1}\mathrm{Mpc}$ are nonlinear at $z=0$, and correspond to halo masses $\sim 10^{14}M_\odot$, $\sim M^{11}M_\odot$ and $\sim 10^8M_\odot$, respectively. 

In Eq.~\eqref{eq:zeta}, the self-similar spikes for small $\chi$ ($|\chi/\chiarm|\ll 1$) are densely packed. For a cosmological scale that is many orders of magnitude above the size of Hubble patch during inflation, these spikes are smoothed out. A few lines of simple algebra give $\zeta_L = A_\zeta/W$ for $|\chi_L/\chiarm| \ll 1$, which is confirmed by the numeric calculation shown in Figure~\ref{fig:zeta}. For a model with $\alpha\gtrsim 10^{-2}$, the smoothing effect significantly differs on various cosmological scales. On linear scales $\gtrsim 10^2h^{-1}\mathrm{Mpc}$, $\zeta$ is almost not modulated by $\chi$ at all. Thus, such models can easily pass the observational test on linear scales. Although, as pointed out by BFHK, it is possible to produce very rare anomalies (such as the CMB cold spot) by extending the $\chi_L$ range to a few $\sigma$'s or by allowing a nontrivial background $\chi/\chiarm$ in the observable universe.

In fact, not much fine-tuning is required to also suppress the modulation on quasi-linear scales all the way down to $\sim 1 h^{-1}\mathrm{Mpc}$. For the model shown in Figure~\ref{fig:zeta} we run simulations for $B=200h^{-1}\mathrm{Mpc}$, $A_{\zeta}=5\times 10^{-3}$ and $\chi_B/\chiarm = 2e^{\lambda W}$, and compute halo mass functions. No noticeable difference of halo mass function in the mass range $\gtrsim 10^{11}M_\odot$ is found between runs with and without modulated preheating.

\begin{figure}
  \includegraphics[width=0.48\textwidth]{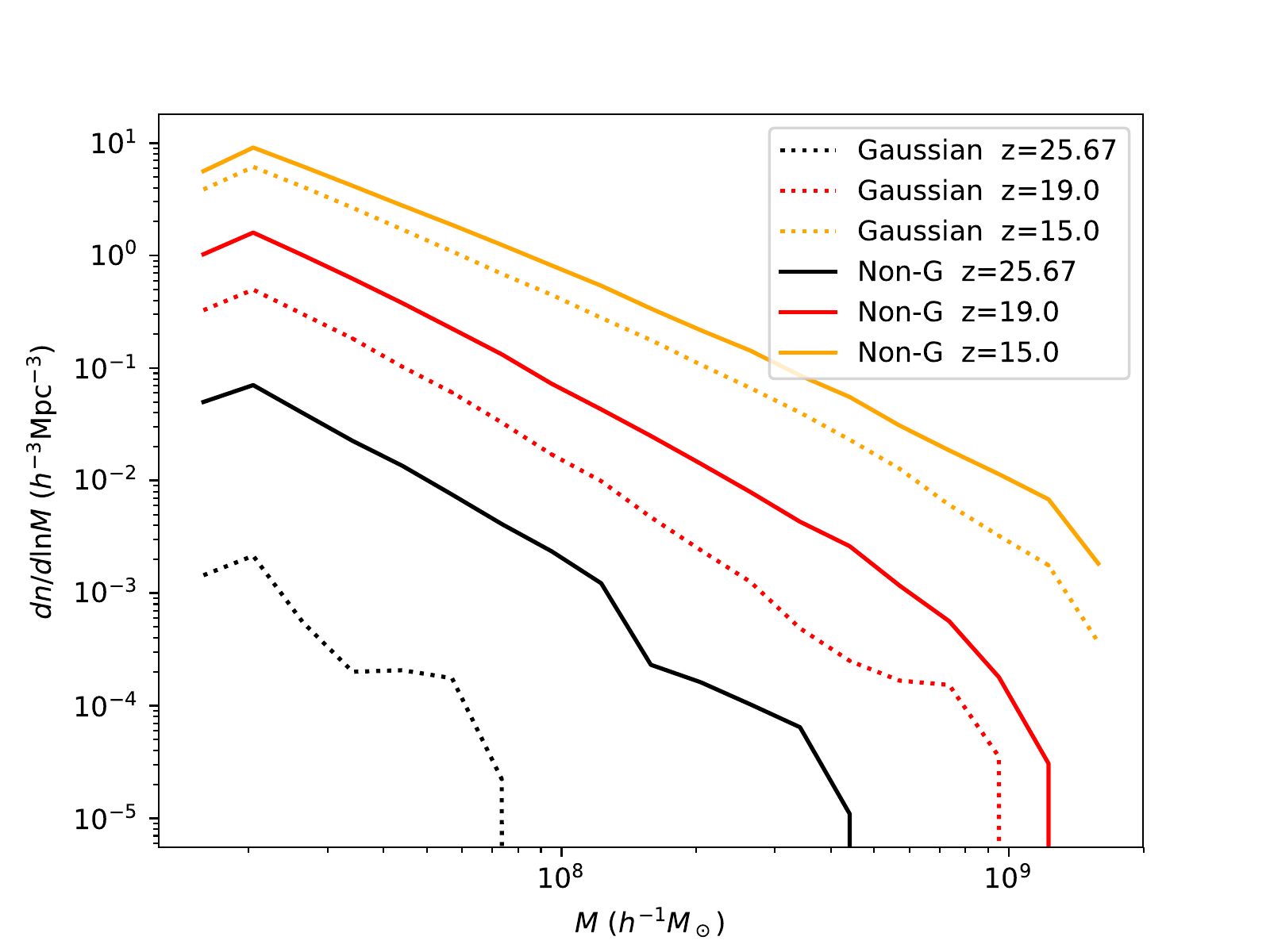}
  \caption{Halo mass functions with (solid lines) and without (dashed lines) modulated preheating.  The parameters for modulated preheating are $\alpha=0.05$, $\lambda=0$, $W=0.5$, and $A_\zeta=0.005$. The results are averaged over 30 COLA simulations with $512^3$ particles and box size $12 h^{-1} \mathrm{Mpc}$.  The background $\chi_B/\chiarm $ is fixed to $4e^{\lambda W}$. \label{fig:hm}}
\end{figure}

The major motivation of this work is to study the abundance of mini-haloes (mass $\gtrsim 10^8M_\odot$)  at very high redshift $z\gtrsim 15$. These haloes are of particular interest, because massive seeds of supermassive black holes (SMBHs) may form in them via direct collapse of primordial gas~\cite{DirectCollapse}. The origin of the increasing number of observed SMBHs at very high redshift is one of the unsolved mysteries in astrophysics. The major theoretical difficulty is their assembly time~\cite{Tanaka09}. In the context of the standard cosmology without modulated preheating, $\gtrsim 10^8M_\odot$ haloes, and hence the direct-collapse SMBH seeds can only form at $z\lesssim 20$. Very efficient accretion, whose viability is yet under debate, is  required in order to explain the growing number of observed bright quasars at $z>6$, which are thought to be powered by $\gtrsim 10^9M_\odot$ SMBHs~\cite{Fan03,Banados18}. With modulated preheating, however, it is possible to produce a significant amount of $\gtrsim 10^8M_\odot$ haloes at $z>20$, which would alleviate the theoretical difficulty in explaining the required SMBH accretion rate. As an example, the evolution of halo mass function at $z>15$ is shown in Figure~\ref{fig:hm}. We continue to use the set of parameters in Figure~\ref{fig:zeta}, which, as we discussed above, has almost no impact on scales $\gtrsim \mathrm{Mpc}$, or equivalently, structures with mass $\gtrsim 10^{11}M_\odot$.

\section{Discussion and Conclusions \label{sec:con}}

We have proposed, and have shown with a concrete example, that the abundance of $\gtrsim 10^8M_\odot$ haloes can be significantly enhanced at very high redshift by viable modulated preheating models. This offers a possible explanation for the observed luminous quasars at $z>6$.

Note that a significant global enhancement of primordial metric fluctuations, even on small scales~$\sim 0.1h^{-1}\mathrm{Mpc}$ is somewhat disfavored by cosmological observations. The key feature of the modulated preheating model, which differs it from many other early-universe models, is that the enhancement of metric fluctuations is typically of intermittent type. For most of the spatial regions where the modulator field $\chi$ is small, the log-uniform spiky responses $\zetah(\chih)$ in Eq.~\eqref{eq:zeta} are smoothed out to a null signal. Only in the (rare) regions where $\chi$ is large, the averaged $\chi\rightarrow \zeta$ mapping becomes non-trivial. In summary, the intermittent feature of metric fluctuations from modulated preheating: (1) naturally explains the rareness of SMBHs at very high redshift; (2) evades tight cosmological constraints on (global) primordial metric fluctuations; (3) makes a general (and falsifiable) prediction that around the high-redshift SMBHs the primordial metric fluctuations tend to be significantly enhanced and thus more small-scale objects should be formed.

Such a phenomenon is indeed quite robust against the variations of parameters, as long as $\alpha$ is kept sufficiently large ($\gtrsim$ a few $\times 10^{-2}$). There is, however, still some tuning in the choice of $\alpha$, which suppresses the modulator power spectrum on very small scales. In other words, the special mass scale $\sim 10^8M_\odot$ is made by tuning down the modulator power spectrum on scales $\lesssim 0.1h^{-1}\mathrm{Mpc}$. Nevertheless, this is not a very fine tuning, and can be naturally realized by an increasing mass of the modulator field towards the end of inflation.

Finally, we would like to point out that a top-hat window function in configuration space does not exactly correspond to a clean cut in Fourier space. In calculations for this article we have ignored such details, because the model itself, namely Eq.~\eqref{eq:zeta}, is a crude phenomenological approximation that may not be worth precision studies. It would be very interesting to construct a concrete example with a full action and confront it with the observations. Another interesting direction is to quantitatively compute the SMBH direct collapse and accretion history with numeric tools developed in the literature. We leave these possibilities for our future works.

\section{Acknowledgment}

I am grateful for Yang Luo's insightful comments on the physics of direct collapse of SMBHs in high-redshift haloes. I also owe thanks to J. Richard Bond, Lev Kofman, Andrei Frolov, Jonathan Braden, and Arttu Rajantie for many intriguing discussions on the topic of modulated preheating.

\end{document}